\documentclass[12pt,reqno]{article}
\usepackage{footmisc}
\usepackage{amsmath,amssymb,amsthm}
\usepackage{CJK}
\usepackage{amsfonts,amssymb}
\usepackage{cite}
\usepackage{longtable}
\usepackage{rotating}
\usepackage{multirow}
\usepackage{setspace}
\usepackage{float}
\usepackage{threeparttable}
\usepackage{graphicx}
\usepackage{indentfirst}
\usepackage{slashbox,pict2e}
\usepackage{multirow}
\usepackage{lineno}
\usepackage{booktabs}
\usepackage[titletoc]{appendix}
\usepackage{mathrsfs}
\usepackage{ulem}
\usepackage[pdfstartview=FitH,%
bookmarksnumbered=true,bookmarksopen=true,%
colorlinks=true,pdfborder=001,citecolor=blue,%
linkcolor=red,urlcolor=blue]{hyperref}
\usepackage[round]{natbib}
\usepackage{booktabs}
\usepackage{psfrag}
\usepackage{amsmath}
\usepackage{amssymb,amsthm, amsfonts}
\usepackage{graphicx}
\usepackage{subfigure}
\usepackage{alphalph}
\usepackage{footmisc}
\usepackage[titletoc]{appendix}
\begin{document}

\title%
{
Using A Partial Differential Equation with Google Mobility Data to Predict COVID-19 in Arizona
 }

\author{
Haiyan Wang \thanks{
Corresponding author(haiyan.wang@asu.edu)
 }\\  
        School of Mathematical and Natural Sciences\\
        Arizona State University\\
        Phoenix, AZ 85069, USA\\
Nao Yamamoto \thanks{
nyamamo6@asu.edu
 }\\
        School of Human Evolution and Social Change\\
        Arizona State University\\
       Tempe, AZ 85287\\
            }        
\maketitle
\begin{abstract}
\noindent{\bf Abstract.}
The outbreak of COVID-19 disrupts the life of many people in the world.  The state of Arizona in the U.S. emerges as one of the country's newest COVID-19 hot spots. Accurate forecasting for COVID-19 cases will help governments to implement necessary measures and convince more people to take personal precautions to combat the virus. It is difficult to accurately predict the COVID-19 cases due to many human factors involved. This paper aims to provide a forecasting model for COVID-19 cases with the help of human activity data from the Google Community Mobility Reports. To achieve this goal, a specific partial differential equation (PDE) is developed and validated with the COVID-19 data from the New York Times at the county level in the state of Arizona in the U.S. The proposed model describes the combined effects of transboundary spread among county clusters in Arizona and human actives on the transmission of COVID-19. The results show that the prediction accuracy of this model is well acceptable (above 94\%). Furthermore, we study the effectiveness of personal precautions such as wearing face masks and practicing social distancing on COVID-19 cases at the local level. The localized analytical results can be used to help to slow the spread of COVID-19 in Arizona.  To the best of our knowledge, this work is the first attempt to apply PDE models on COVID-19 prediction with the Google Community Mobility Reports. 

\end{abstract}
\section{Introduction}


The outbreak of COVID-19 is disrupting the personal lives of people around the world in a variety of ways and has made profound impacts on the world socially and economically.  As the number of confirmed COVID-19 cases in the U.S. continues to rise in early 2020, many states declared states of emergency and shutdown orders or stay-at-home orders were issued to slow the spread of COVID-19. Many schools, workplaces, and public gathering spaces across the U.S. were closed for an extended time. In June 2020, as states gradually reopen the economy, the number of COVID-19 cases starts to increase again in many states.
 
In June 2020, Arizona emerged as one of the country's newest COVID-19 hot spots with an alarming resurgence of COVID-19 after the stay-at-home order was lifted in the middle of May. The number of people hospitalized is also climbing. On June 15, Arizona reported 2,497 new COVID-19 cases, a new all-time high record, up from the previous highest number of cases of 2,140 cases in a single day on June 12. Previously, Arizona experienced the first rapid rise in the number of confirmed cases of COVID-19 infection on March 16 with a total of 1,157 confirmed cases including 20 deaths, according to the Arizona Department of Health Services (\cite{ADHS2020}). The spike of the new COVID-19 case is believed to result from the fact that Arizona took a more aggressive approach to reopen the state in the middle of May. In addition, there is also no state-level requirement for everyone to wear face masks. Many Arizonans have different views on practice social distancing which is the only way to prevent infections in the absence of a vaccine. Although under Arizona's reopening plan, businesses are only advised to follow federal guidance on social distancing, the trade-offs between health and the economy have forced many businesses to choose not fully implementing the social distancing requirements during the COVID-19 pandemic. Therefore, quantifying the impact of social distancing on minimizing the epidemic impact can make governments and convince more people to better understand the significance of social distancing and personal precautions. In the present paper, we examine the COVID-19 situation in the state of Arizona in the United States and predict the number of cases in Arizona after the reopening of the economy. 

There have been plenty of mathematical models to describe the spread of infectious diseases. To reply to the current COVID-19 pandemic, many modeling studies using ordinary differential equations (ODE) have proposed for predicting the COVID-19 cases (e. g. \cite{He2020MBE,Li2020MBE,Wang2020MBE}). The classical susceptible infectious recovered model (SIR)  (\cite{Huang2020BullWHO}) and susceptible exposed infectious recovered model (SEIR) (\cite{Lai2020Nature, Omori2020MedRxiv, Yang2020JTD}) are the most widely adopted ones for characterizing the outbreak of COVID-19. The extension of the classical SEIR model with age-stratified model (\cite{Prem2020LancetPH}) and meta-population model (\cite{Pujari2020medRxiv}) were also introduced.

In this paper, we present a spatio-temporal model, specifically, a partial differential equation (PDE) model, of COVID-19 based on county-level clusters.  The proposed model describes the combined effects of transboundary spread among county clusters in Arizona and human activities on the transmission of COVID-19. This is, in particular, important for Arizona as well as the other states in the U.S. because new COVID-19 cases in the U.S. are still continuing to head upward after the lockdown ended while most countries had peaked their new cases after lockdowns. One of the factors contributing to the continued spike of the U.S. cases is that personal precautions are not consistent across the U.S. once the lockdown ends. In Arizona, the face mask requirement is implemented at the local government level. A number of counties in Arizona do not require face masks. We divided the counties in Arizona into three regions. The three regions represent different geographical and social characteristics in terms of the spread of COVID-19. These disparities, which reflect differing regional levels of risk, can be better modeled by a spatio-temporal model. As a result, the characteristics of the PDE model make it suitable for COVID-19 prediction. The localized results of the spatio-temporal model provide more useful information for the local governments to closely monitor new COVID-19 clusters and quickly reinstating lockdowns at the local level when epidemic spikes. 
 
Many spatio-temporal models characterize infectious diseases by PDEs, describing the dynamics of susceptible, infected, and recovered populations (e.g. \cite{Brauer2019book, holmes1994ecology, Wang2020mdpi, Zhu2017MBE}). Our previous work (\cite{Haiyan2019book,wang2016regional}, \cite{Wang2020mdpi}) applies PDE models to make a regional level of influenza with geo-tagged data.  The PDE model we develop in the present work focuses on the spread of COVID-19 and incorporating social distancing factors.  While there is a rich literature on the application of PDE to modeling the spatial spread of infectious diseases, to the best of our knowledge, this work is the first attempt to apply PDE models on COVID-19 prediction with real-data validation. 

In addition, our model is different from the existing models in that we include the data from the Google Community Mobility Report  (\cite{Google2020Mobility}) which reflects the effects of human activities. To help to combat the spread of COVID-19, Google releases the COVID-19 Community Mobility Reports which provide daily, county-level aggregated data on time spent at different categories of activities, compared with a baseline period before the epidemic.  These activities include retail and recreation, groceries and pharmacies, parks, transit stations, workplaces, and residential areas. Mobility trends obtained from location history are dynamic in time and reflect real-time changes in social behavior, making them a crucial factor in analyzing COVID-19 spread and its countermeasure. There are several ODE and statistical models which utilize the COVID-19 Community Mobility data~(\cite{picchiotti2020arxiv,abouk2020arxiv,voko2020geroscience}). To the best of our knowledge, this is the first PDE model incorporating COVID-19 Community Mobility data to predict the number of COVID-19 cases. 


\section{Data Description}
\label{data}
\begin{figure}[!h]
\centering
\includegraphics[width=0.5\textwidth]{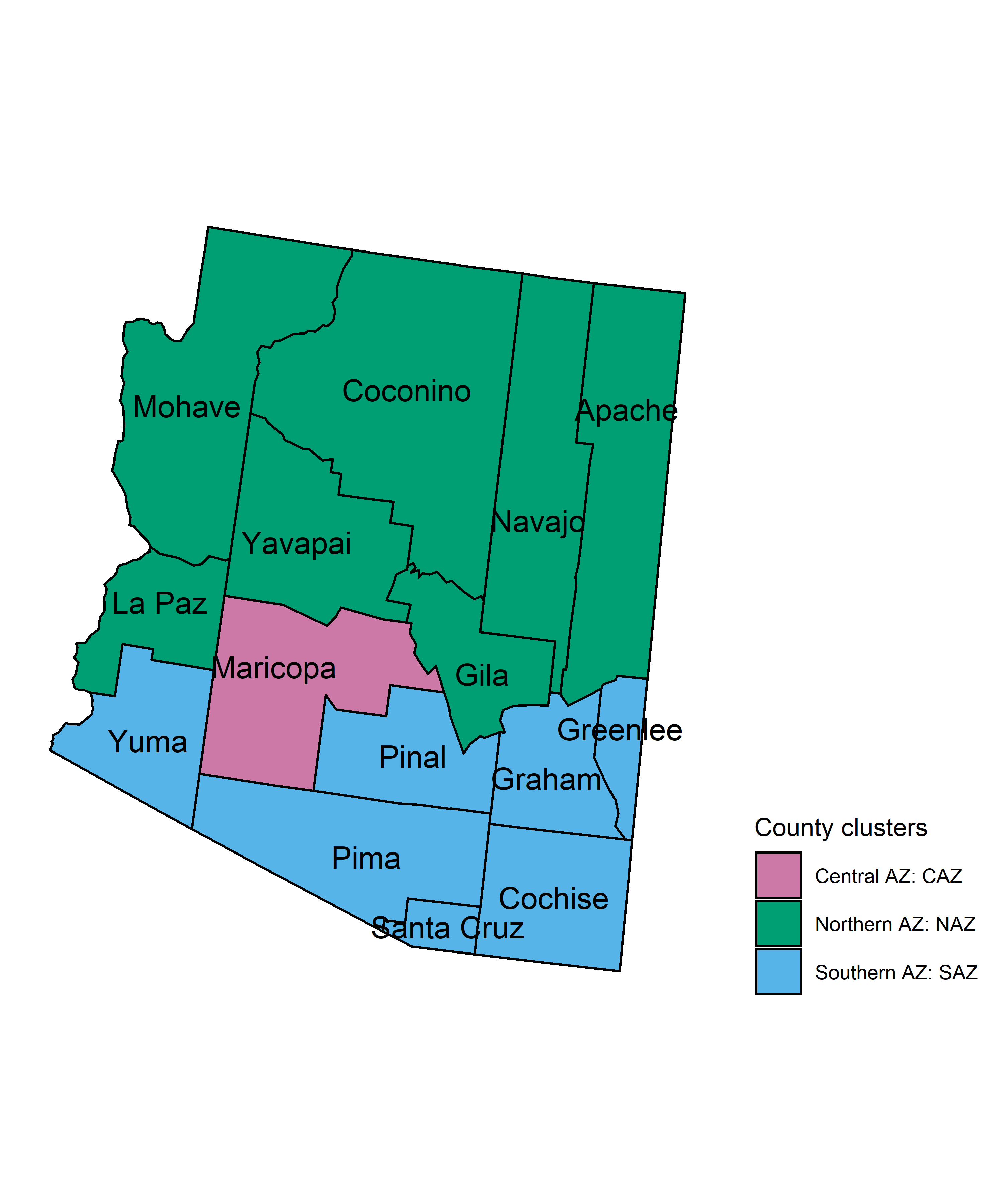}
\caption{Three regions in Arizona
  }
\label{regions}
\end{figure}
There are 15 counties in the U.S. state of Arizona. In general, Arizona is divided into Northern, Central, and Southern Arizona. Although the boundaries of the regions are not well-defined, as we aim to predict COVID-19 cases in Arizona, we group the counties of Arizona into the three regions in Fig. \ref{regions}. The three regions have better balanced COVID-19 cases because Maricopa County, where the city of Phoenix is located, has about 60\% of the population in Arizona. More importantly, they represent different geographical and social characteristics in terms of the spread of COVID-19. Much of the Northern region is National Forest Service land, parkland, and includes part of Navajo reservations. For the per-capita COVID-19 infection rate in the US, the Navajo Nation, which spans parts of Arizona, New Mexico, and Utah, is greater than that of New York City in May 2020. The Southern Arizona includes the city of Tucson and several other small cities close to the border of the U.S. and Mexico.  In the following, we will use NAZ (Region 1), CAZ (Region 2), and SAZ (Region 3) to denote the clusters of the Arizona counties.

\begin{itemize}
\item  NAZ (Region 1)=[Mohave County, Coconino County, Yavapai County, La Paz County, Gila County, Navajo County, Apache County].
\item  CAZ (Region 2)=[Maricopa County].
\item  SAZ (Region 3)=[Yuma County, Pima County, Pinal County, Graham County, Santa Cruz County, Greenlee County, Cochise County].  
\end{itemize}
 
We compute the daily new cases of each region by adding the new COVID-19 cases of all counties belonging to a region. We use the COVID-19 data from the New York Times at the state and the county level over time. The New York Times is compiling the time series data from state and local governments and health departments in an attempt to provide a complete record of the ongoing outbreak. The data can be downloaded from https://github.com/nytimes/covid-19-data and https://www.nytimes.com/article/coronavirus-county-data-us.html. 
 
In addition, we incorporate the Google Community Mobility Report into our model. These Community Mobility Reports provide insights into how people's social behaviors have been changing in response to policies aimed at combating COVID-19. The reports provide the changes in movement trends compared to baselines overtime at the U.S. county level, across different categories of activities. These activities include retail and recreation, groceries and pharmacies, parks, transit stations, workplaces, and residential. The relevant data can be downloaded from https://www.google.com/covid19/mobility/. Then we compute the daily changes of each region by adding the changes of all counties belonging to a region.

We create two time series from the Google Community Mobility Report. We accumulate the data for retail \& recreation, groceries \& pharmacies, parks, transit stations, workplaces, which are believed to increase the number of COVID-19 cases. On the other hand, the data from Google mobility for residential activities describes stay-at-home activities that are believed to prevent COVID-19 epidemics. In some cases, the increase of residential activities may result from events (such as parties) that could promote COVID-19 cases. However, the Google data only indicates residential activities compared to baselines. In our model, we use a Michaelis–Menten function to limit the effect of residential activities.  Because it may take about two weeks for the effect of the activities to take place, we subtract the time $t$ in our PDE model by 14 to reflect the incubation period of COVID-19 in most cases.

\section{PDE Model}
\label{models}
\begin{figure}[!h]
\centering
\includegraphics[page=1,scale=0.7, trim=7cm 7cm 7cm 7cm, clip=true ]{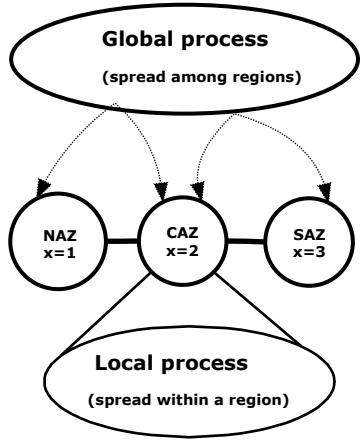}
\caption{Embedding of three regions into the $x$-axis and two spreading processes}
\label{embed}
\end{figure}

In this section, we introduce a specific PDE model to characterize the spatio-temporal dynamics of COVID-19 cases. To apply a PDE model to the interaction of the dynamics of COVID-19 cases, one must embed the three regions NAZ, CAZ, SAZ into  Euclidean space and arrange them in a meaningful order.  In this paper, the three regions are mapped onto a line where the locations of NAZ, CAZ, SAZ are 1, 2, 3 respectively. The arrangement will make the three regions stay as close as possible for ensuring that the continuous model can capture the spread of COVID-19 cases between the three regions.  Projecting the three regions, geographically located in the north-south direction of Arizona, onto the x-axis of the Cartesian coordinates, as shown in Fig. \ref{embed}. We project the three regions as the order of geographical north-south direction instead of other orders mainly due to the following fact: Maricopa county has much more population than any other county in Arizona. It is more likely that COVID-19 spreads from the metro Phoenix area to both northern and southern counties, at least at the beginning of 2020. 

Having embedded the three regions to the Euclidean space, we will develop the spatio-temporal model for the spread of COVID-19 cases according to the balance law: the rate at which a given quantity changes in a given domain must equal the rate at which it flows across its
boundary plus the rate at which it is created, or destroyed, within the domain. The PDE model can be conceptually divided into two processes: an internal process within each region and an external process among regions. Similar derivation for the PDE model has been used in our previous work for PDE models for information diffusion in online social networks in \cite{Haiyan2019book}.

Let $C(x,t)$ represent the COVID-19 cases in Arizona region $x$ at a given time $t$. The changing rate of $C(x,t)$ depends on two processes as in Fig.~\ref{embed}:
\begin{itemize}
\item  Global process: the social interactions of people between regions, such as traveling and commuting between regions, that contribute to the spread of COVID-19;
\item Local process:  in each Arizona region, people become newly infected through social interactions with infected people within a region; and people may take personal precautions to reduce and mitigate COVID-19 spread. 
\end{itemize}

Combining the two processes, the dynamics of COVID-19 cases can be captured by Eq.~(\ref{new_model}).
\begin{eqnarray}
\label{new_model}
\left\{\begin{array}{ll}
&\frac{\partial C(x,t)}{\partial t}=\frac{\partial }{\partial x}\big(d(x)\frac{\partial C(x,t)}{\partial x}\big)+r(t)l(x)a(x,t-14)C(x,t)-c\frac{h(x,t-14)C(x,t)}{k+C(x,t)},
\\
&C(x,1)=\psi(x),1<x<3,
\\
&\frac{\partial C}{\partial x}(1,t)=\frac{\partial C}{\partial x}(3,t)=0,t>1,
\end{array} \right.
\end{eqnarray}
where
\begin{itemize}
          \item The term $\frac{\partial}{\partial x} \left(d(x)\frac{\partial C(x,t)}{\partial x}\right)$ denotes the spread of COVID-19 cases between different Arizona regions, where  $d(x)$ measures how fast COVID-19 spreads across different Arizona regions. In epidemiology~(\cite{Brauer2019book} and \cite{natr2002murray}), the term $\frac{\partial}{\partial x} \left(d(x)\frac{\partial C(x,t)}{\partial x}\right)$ has been widely used for describing the spatial spread of infectious diseases. Here we assume $d(x)$ to be constant, i.e., $d(x)\equiv d>0$.
   
       \item $r(t)l(x)C(x,t)l(x)a(x,t-14)$ represents the new COVID-19 cases from a local Arizona region at location $x$ and time $t$. This type of function is widely used to describe the growth of bacteria, tumors, or social information over time~(\cite{natr2002murray}).  
          \begin{itemize}
         \item[$\ast$] The function $r(t)>0$ represents the growth rate of COVID-19 cases at time $t$ for all Arizona regions. For simplicity, we assume that $r(t)$ increases with time $t$ as the COVID-19 cases increase. Therefore, we choose $r(t)=g(b_1+b_2 t)$ and $g(u)=1/(1+exp(-u))$ to describe the pattern with parameters $b_1>0, b_2>0$ to be determined by the collected COVID-19 data.
         \item[$\ast$]The location function $l(x)$ describes the spatial heterogeneity of COVID-19, which depicts different infection rates in the three Arizona regions. The function $l(x)$ is built through a cubic spline interpolation, which satisfies $l(x_i)\equiv l_i, i=1,2,3,$ where $x_i$ represents the location of Arizona region $i$.  $l(x)$ is determined by the collected COVID-19 data.
         \item[$\ast$] $a(x,t-14)$ is accumulated data from Google mobility for retail \& recreation, groceries \& pharmacies, parks, transit stations, workplaces, which are believed to increase the COVID-19 cases.  $t-14$ again reflects the incubation period of COVID-19 in most cases.  
        \end{itemize}

    \item The function $c \frac{h(x,t-14)C(x,t)}{k+C(x,t)}$ is the rate of decrease of COVID-19 cases due to human efforts. During the period of the epidemic, Arizona people are advised to take COVID-19 precaution actions, such as wearing masks, staying six feet apart, or staying at home to reduce the contact rate. $c\frac{h(x,t-14)C(x,t)}{k+C(x,t)}$ to describe the potential reduction in COVID-19. 
          \begin{itemize}
          \item[$\ast$] In most cases, the data from Google mobility for residential activities are believed to prevent the spread of COVID-19. In some cases, they may result from events (such as home parties) that could promote COVID-19 cases.  Here we use Michaelis–Menten function to limit the effect of home activities in the spread of COVID-19. 
           \item[$\ast$]  $c>0$ represents the maximum reduction rate of COVID-19 cases due to government measures and personal precautions to maintain social distancing including wearing masks, staying six feet apart. $k$ is the number of COVID-19 cases at which the reduction rate is $\frac{1}{2} c $. By adjusting the parameters, this model allows us to examine the effect of social distancing to prevent COVID-19 cases.
           \item[$\ast$] $h(x,t-14)$ describes home activities which are believed to prevent COVID-19 epidemics.  $h(x,t-14)$ is accumulated data from Google mobility for home activities.  $t-14$ is because it takes two weeks for the effect of the activity to take place.  
          \end{itemize}

        \item $a(x,t-14), h(x,t-14)$ are determined by Google Mobility data with the interpolation function of Matlab.
        \item $d,k,c$ and parameters of $r(t), l(x),$ are determined by the known historical data of the COVID-19 cases with the best fitting function of Matlab. 
       
       \item Neumann boundary condition $\frac{\partial C}{\partial x}(1,t)=\frac{\partial C}{\partial x}(3,t)=0, \quad t>1$ is applied in~ (\cite{natr2002murray}). For simplicity, we count the cases imported from neighbor states as local Arizona cases and assume that no COVID-19 spreads across the boundaries at $ x=1, 3$.   
       
       \item Initial function $C(x,1)=\psi(x)$ describes the initial states of COVID-19 in every Arizona region, which can be constructed from the historical data of COVID-19 cases by cubic spline interpolation.
\end{itemize}

\section{Model Prediction}
\label{prediction}

\begin{figure}[!h]
\centering
\subfigure[]{\includegraphics[width=3.2in, trim=3cm 8cm 3cm 8cm, clip=true]{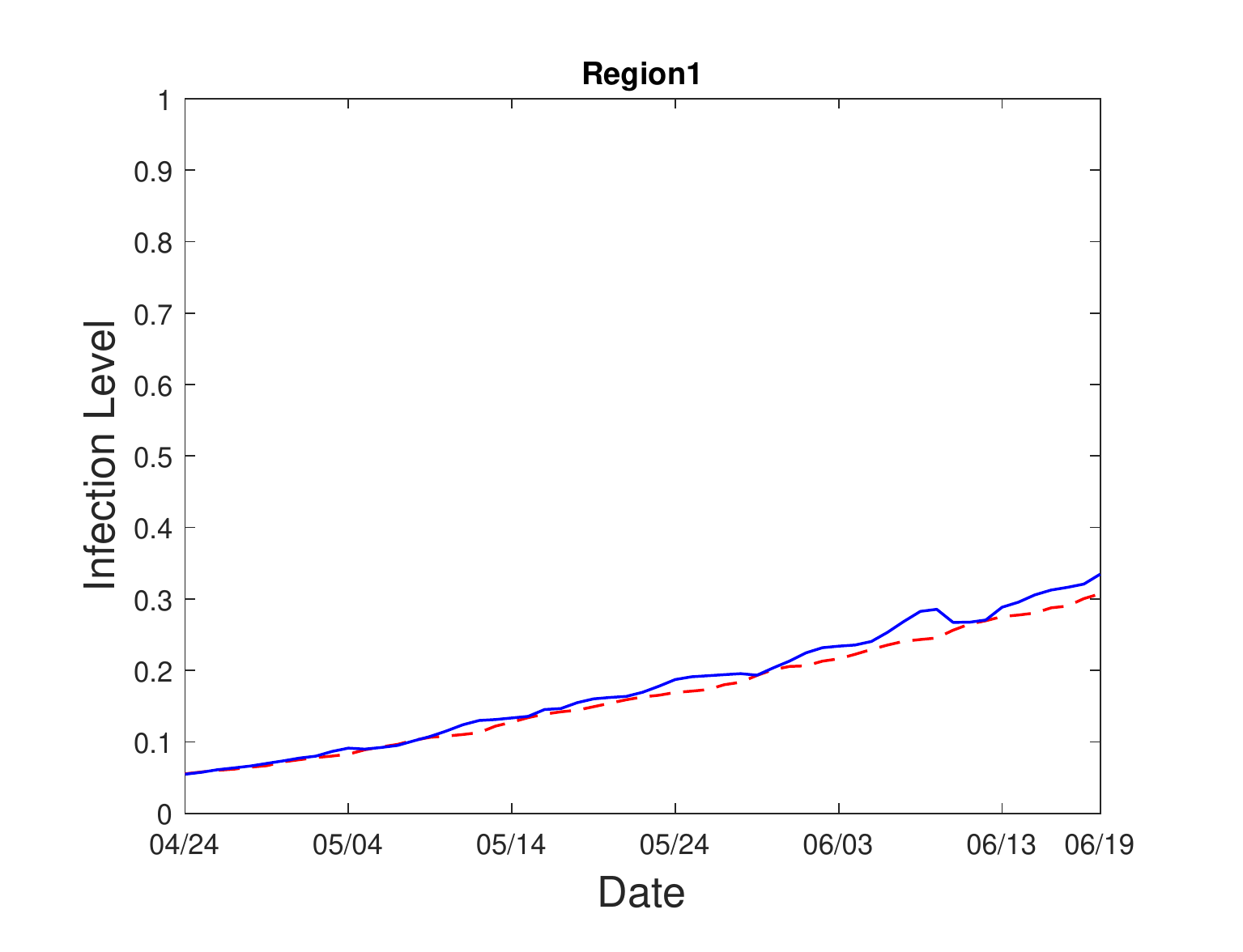}}\
\subfigure[]{\includegraphics[width=3.2in, trim=3cm 8cm 3cm 8cm, clip=true]{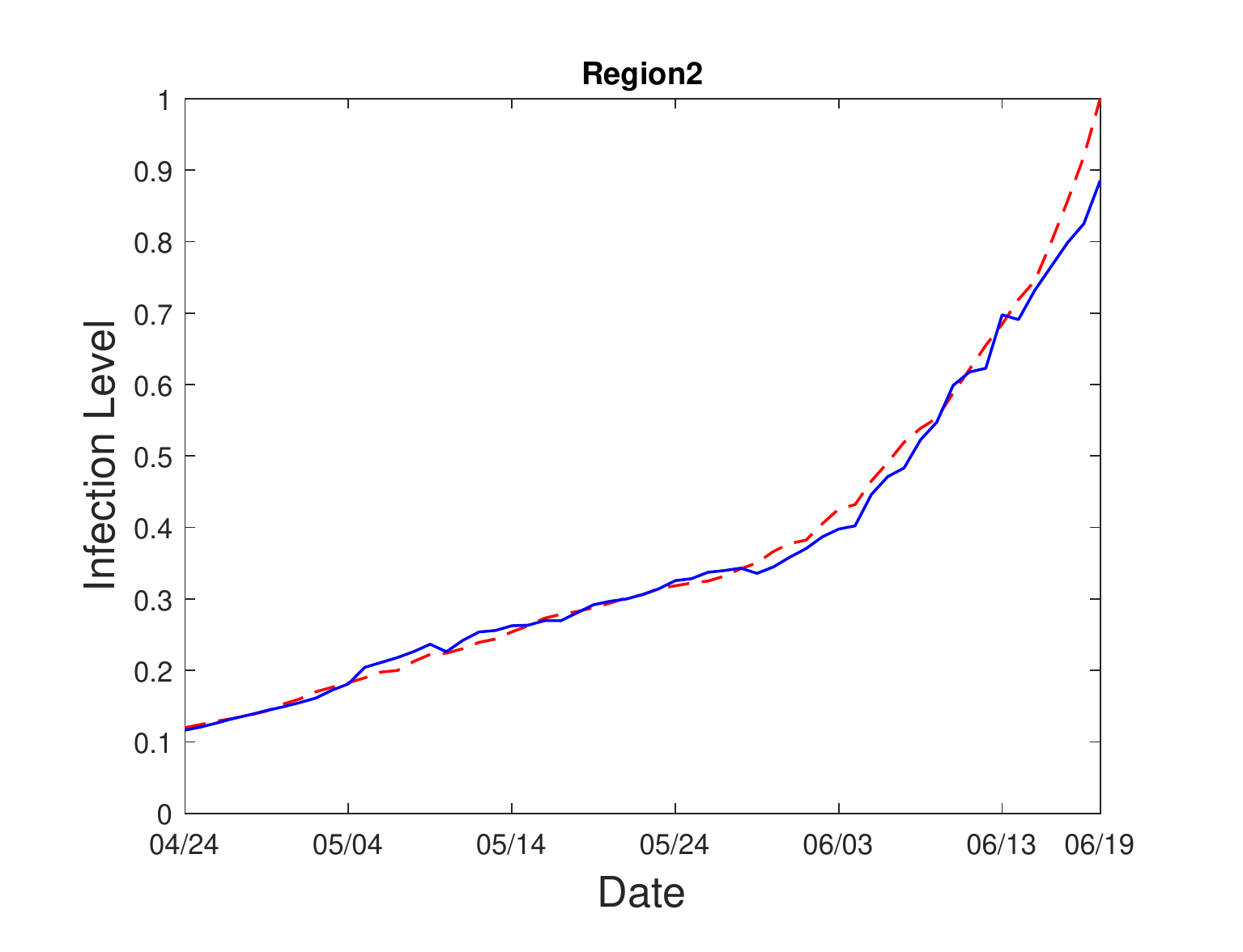}}\
\subfigure[]{\includegraphics[width=3.2in, trim=3cm 8cm 3cm 8cm, clip=true]{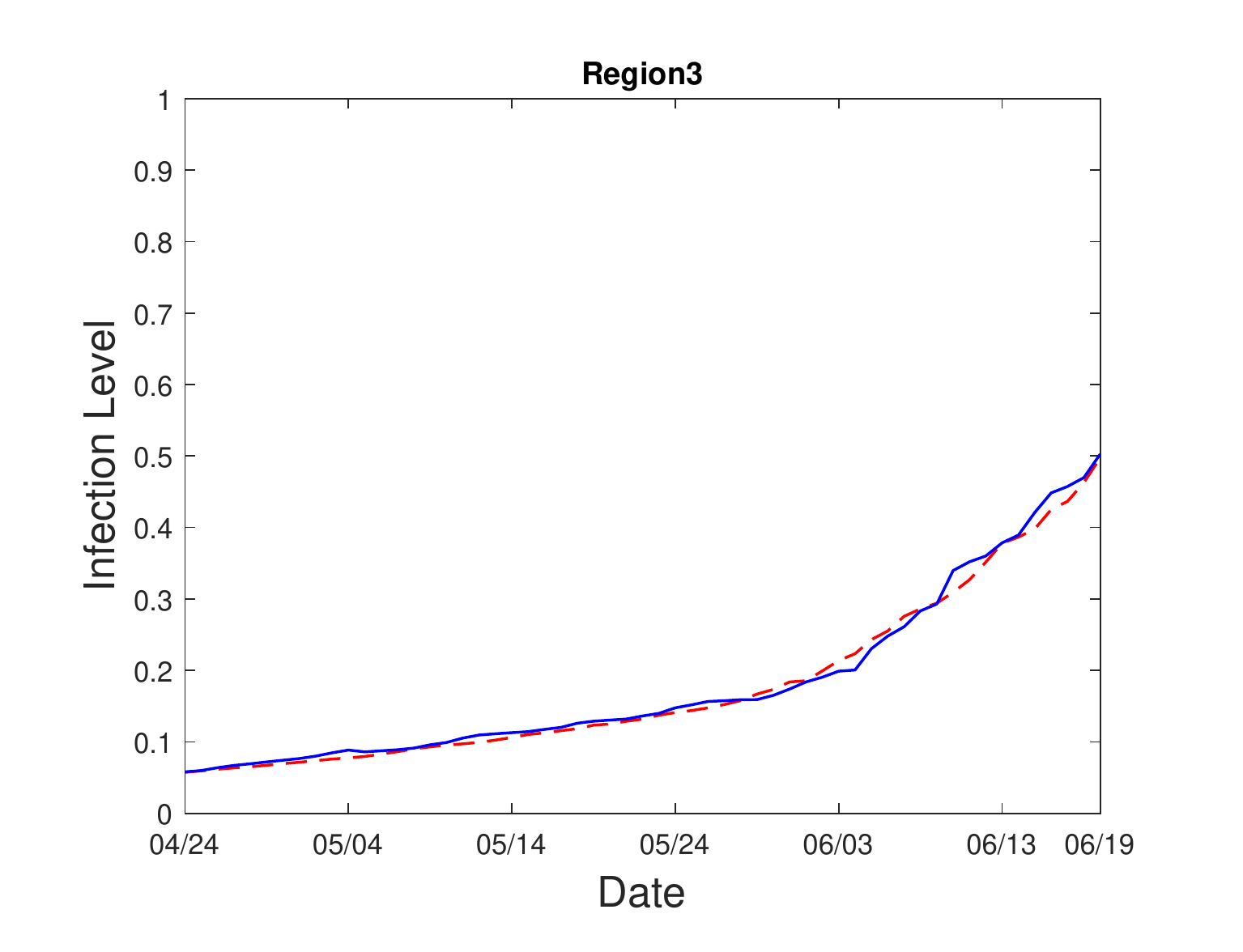}}\
\caption{Predictions of COVID-19 cases in Arizona from April 24, 2020, to June 19, 2020, for three regions. The blue lines represent the predicted COVID-19 cases and red lines represent the actual COVID-19 cases.}\label{Predictions}
\end{figure}

\begin{figure}[!h]
\centering
\subfigure[]{\includegraphics[width=3.1in, trim=3cm 8cm 1cm 6cm, clip=true]{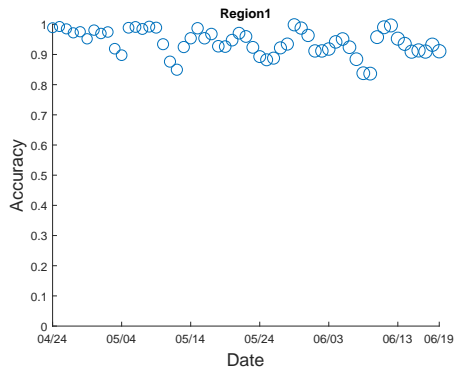}}\
\subfigure[]{\includegraphics[width=3.1in, trim=3cm 8cm 1cm 6cm, clip=true]{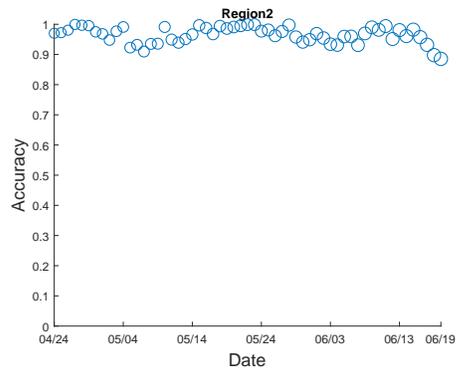}}\
\subfigure[]{\includegraphics[width=3.1in, trim=3cm 8cm 1cm 6cm, clip=true]{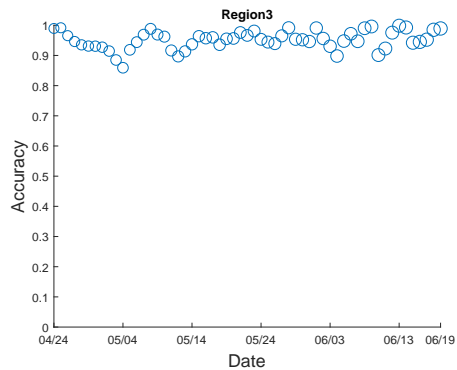}}\
\caption{The relative accuracy in each region: NAZ (Region 1), CAZ (Region 2), and SAZ (Region 3) from April 24, 2020 to June 19, 2020.}\label{RA}
\end{figure}

The basic mathematical properties of the proposed PDE model in Eq.~(\ref{new_model}) such as existence and uniqueness can be established from the standard theorems for parabolic PDEs in~\cite{friedman2008partial}. Below, we evaluate the robustness of our PDE-based predictive model and validate if the model has acceptable prediction performance with the COVID-19 case report in Arizona. The procedure of predictive modeling for the COVID-19 case in Arizona is summarized as follows:
\begin{itemize}

\item \textbf{Prediction Procedure:} In the prediction process of the research time period, the model parameters are to be determined by the historical aggregated data of COVID-19 cases in the three regions under the same structured PDE. In the current experiment, we predict the COVID-19 cases three days ahead. In order to forecast the COVID-19 of a given day, we first train the parameters of the PDE model and then solve the PDE model for prediction. For example, days 1--7, 2--8, \ldots, are used as the training data, and we predict the number of COVID-19 cases for the following days 10, 11, \ldots, respectively. 

The process of performing the prediction can be divided into two major processes. We first use an optimization method to fit parameters in the PDE model with historical data of COVID-19 cases. In essence, this is a multi-parameter inverse problem of parabolic equations. We integrate the local and global methods to search for the best fitting parameters.  In the present work for COVID-19 modeling problems, we take a hybrid approach: first, a tensor train global optimization (\cite{oseledets2011tensor}) is used to explore the parameter space thus to locate the starting points and then Nelder-Mead simplex local optimization method (\cite{lagarias1998convergence}) is used to search the local optimization of COVID-19 modeling problem. The Nelder-Mead simplex method is implemented in the fminsearch function in Matlab. Once the model parameters are determined, we use the fourth-order Runge-Kutta method to solve the PDE for one-step forward prediction numerically.
  
 \item \textbf{Accuracy Measurement:}
We need to quantify the accuracy of model prediction of the COVID-19 cases by comparing the predicted COVID-19 cases with the observed aggregated COVID-19 cases for three regions, which are the ground truth. The mean absolute percentage error $$1-\left|\frac{x_{real}-x_{predict}}{x_{real}}\right|$$ is applied to measure the prediction accuracy, $x_{real}$ is the observed COVID-19 cases at every data collection time point and $x_{predict}$ is the predicted cases. 

\end{itemize}

Fig.~\ref{Predictions} illustrates predictions of COVID-19 cases in Arizona from April 24, 2020, to June 19, 2020, for three regions. The period covers the period of the end of stay-home order when the COVID-19 case started to increase quickly. The blue lines represent the predicted COVID-19 cases and red lines represent the actual COVID-19 cases.  Note that we normalize the data to be between $[0,1]$. Parameters in each prediction step are different, here we just provide the parameters in the last prediction for June 19.  The parameters for prediction on June 19 are: $d=0.013556543719$, $k=4.655312010389$, $b_1=0.503402937243188$, $b_2=0.236417906835504$, $c=0.821575737918722$ and $l_1=0.0826411111659247$, $l_2=0.347970638795282$ and $l_3=0.28107732159028$. Fig.~\ref{RA} illustrates the prediction accuracy in NAZ, CAZ, and  SAZ every day from April 24, 2020, to June 19, 2020. The Arizona stay-home order started on March 31, 2020, and ended on May 15, 2020. The average relative accuracy of three regions with three days prediction are 94\%, 97\%, and 95\% respectively. Clearly, the PDE-based prediction model has a good short-term prediction ability during active COVID-19 epidemics. 


\section{Effects of Social Distancing and Face Covering}

\begin{figure}[!h]
\centering
\subfigure[]{\includegraphics[width=3.2in, trim=3cm 8cm 3cm 8cm, clip=true]{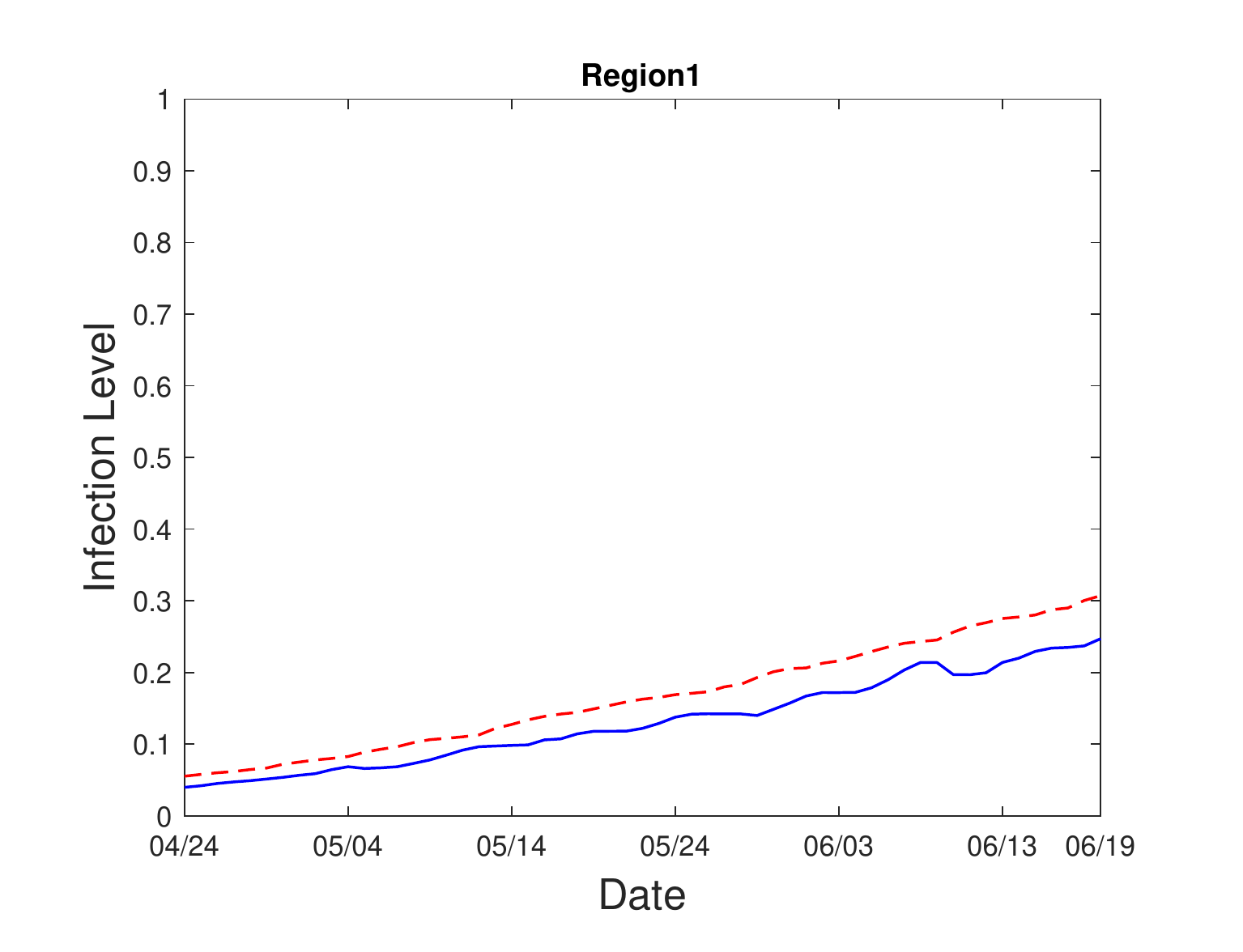}}\
\subfigure[]{\includegraphics[width=3.2in, trim=3cm 8cm 3cm 8cm, clip=true]{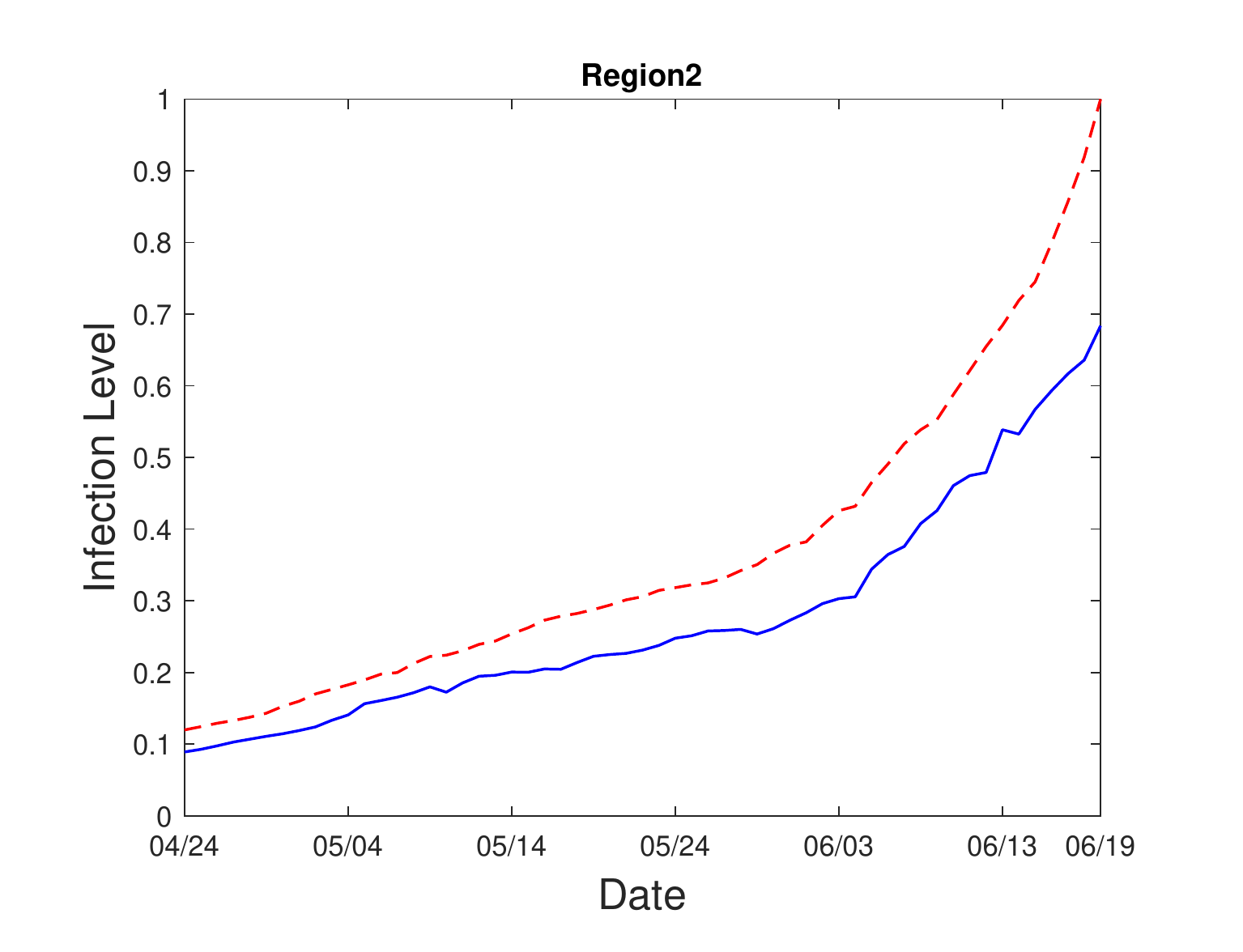}}\
\subfigure[]{\includegraphics[width=3.2in, trim=3cm 8cm 3cm 8cm, clip=true]{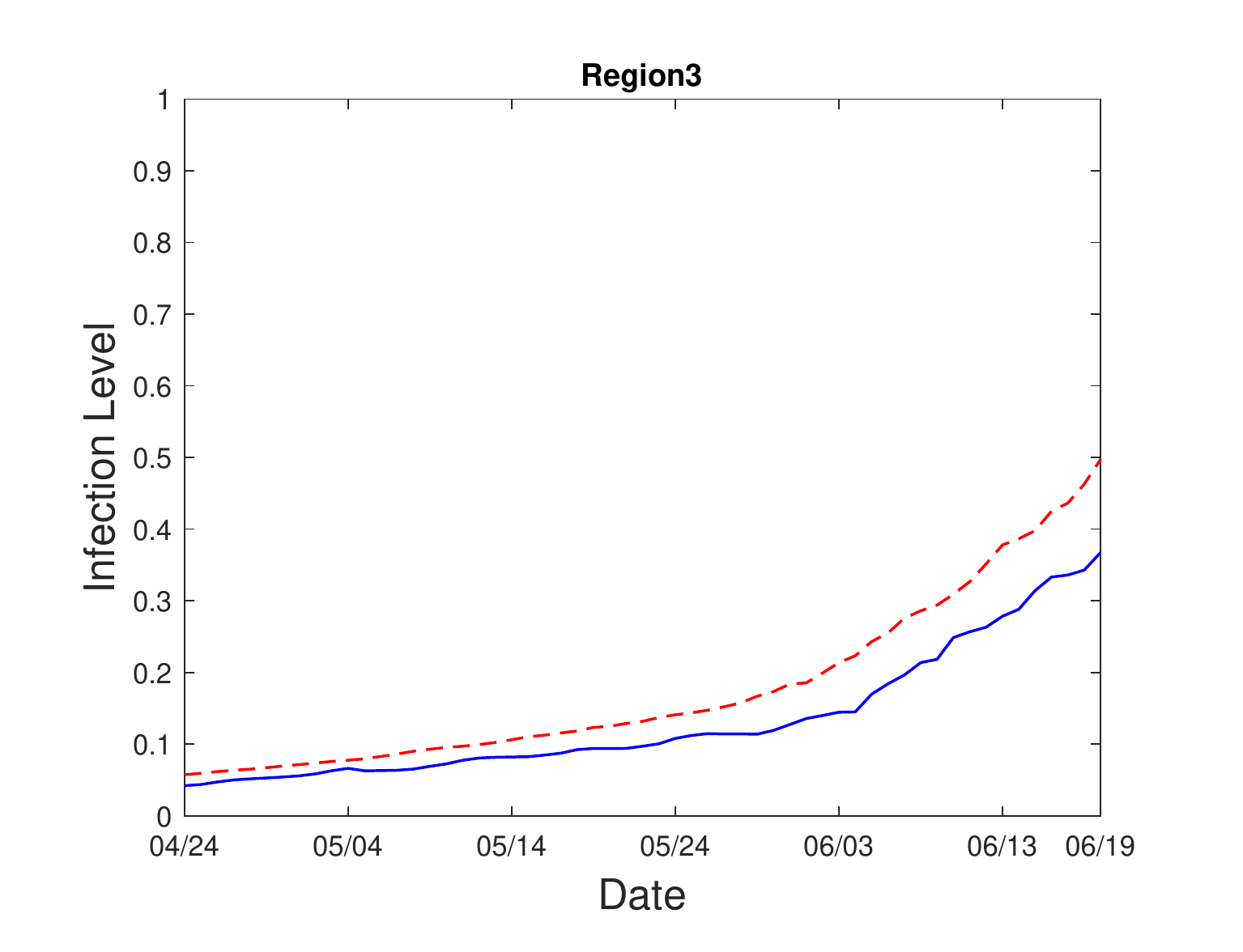}}\
\caption{Effects of preventive measures on the growth of COVID-19 cases in Arizona from April 24, 2020, to June 19, 2020. The red lines represent the actual COVID-19 cases.  (a), (b), and (c) capture the predicted COVID-19 cases (blue lines) under additional social distancing and face covering to reduce COVID-19 for Region 1, 2 and 3. }\label{PolicyWhole}
\end{figure}

\begin{figure}[!h]
\centering
\includegraphics[width=0.5\textwidth]{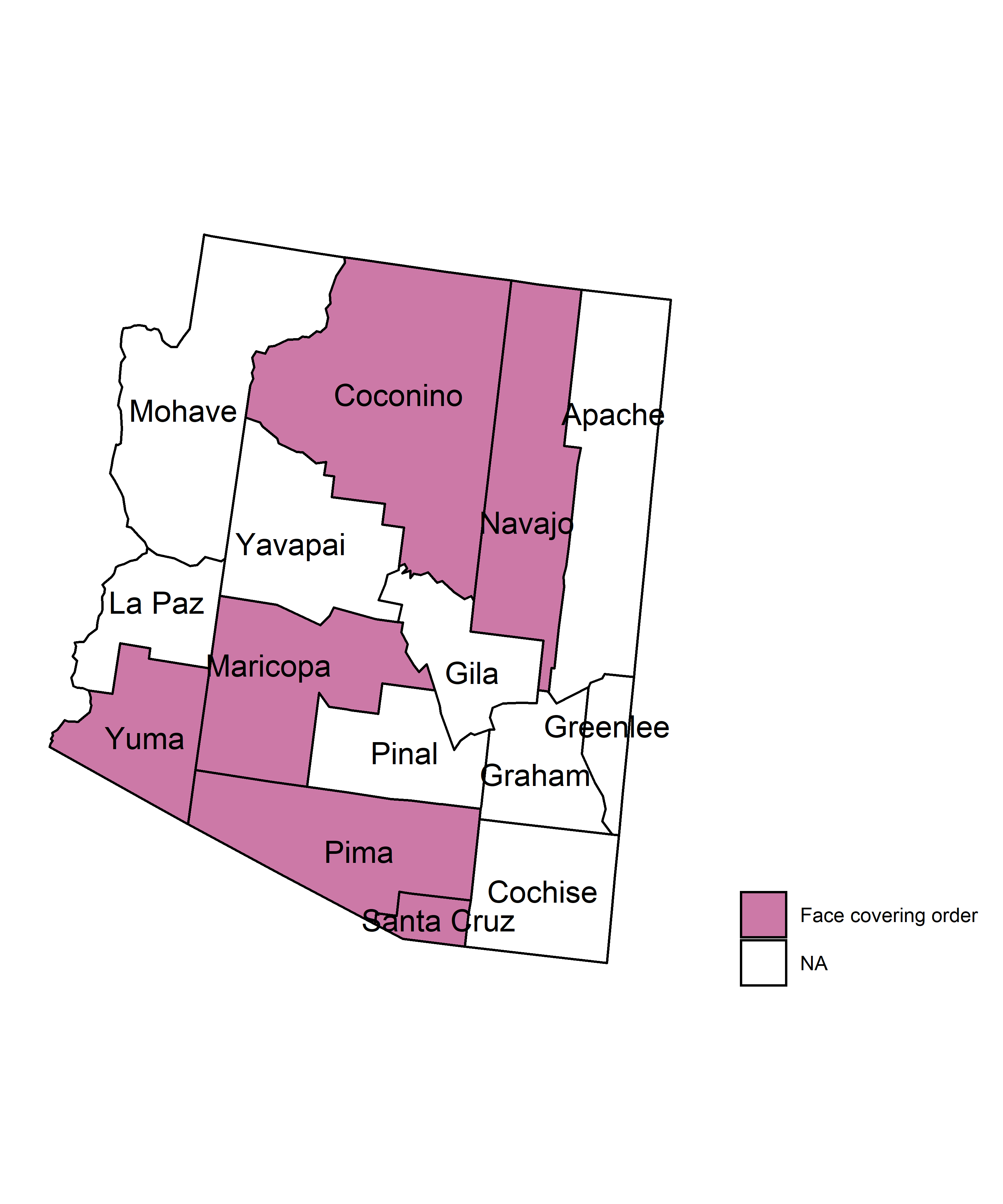}
\caption{Arizona counties with face covering order}
\label{mask}
\end{figure}

To demonstrate the effectiveness of intervention measures such as face masks and social distancing on decreasing COVID-19 cases, we predict the COVID-19 cases with  an increased value of $c$ which measures the effect of the stay-at-home-order and social distancing.  We do not have specific data from Google Community Mobility to incorporate our model for the effectiveness of social distancing and face mask. In general, it is believed that face masks and proper social distancing could significantly reduce COVID-19 cases. Specifically, based on each group of model parameters obtained in the above prediction process, we retain them unchanged in Eq.~(\ref{new_model}) except doubling the value of $c$ (the rate of effectiveness of social distancing).  The doubled value of $c$ suggests that, to reduce COVID-19 cases, people in Arizona take the extra effort of practicing social distancing and wearing face covering even after the lift of the stay-at-home-order.
 
Fig.~\ref{PolicyWhole} illustrates the COVID-19 cases with possible interventions, where the blue lines represent the predicted COVID-19 cases from April 24, 2020, to June 19, 2020, and the red lines represent the actual COVID-19 cases.   Fig.~\ref{Predictions}  can serve as baseline references. Fig.~\ref{PolicyWhole}  (a), (b), (c) captures the predicted COVID-19 cases (blue lines) under double measurements to reduce COVID-19 for Region 1, 2 and 3 respectively.   Fig.~\ref{PolicyWhole} shows that if stricter precaution measures are taken, the COVID-19 cases will be controlled to some extent. As a result, our analysis confirms the positive short-term effects of precaution measures on the spread of COVID-19. 

Unfortunately, there is no mandatory requirement of face-covering at the state level.  About one month after the stay-at-home-order expired on May 15, local governments established their own health and safety measures such as face mask mandates in an effort to slow the spread of COVID-19. As a result, the state keeps breaking a record of daily cases and hospitalizations suspected to be related to COVID-19.  CAZ (Maricopa county) continues to have the highest number of cases in Arizona. On June 22, multiple officials from cities across the state began implementing requirements for face coverings in public. There have been five counties including Maricopa County that have a mask requirement in place as shown in Fig.~\ref{mask} as of June 29, 2020.  Many cities and towns across metro Phoenix already had enacted their own face-covering-mandates in public earlier but Maricopa County (CAZ) took the lead in sending a clear message to every resident in the county to wear a mask. We hope the COVID-19 cases will begin to go downward in the near future.

\section{Discussion}
\label{discussions}
In this paper, the proposed PDE model is built on three regions in Arizona for the prediction of COVID-19 cases. The PDE model is capable of capturing the effects of various geographical and social factors. To ensure high accuracy predictions of COVID-19 cases, daily cases of COVID-19, and the effects of social distancing are included in the prediction model.  In addition, Google Community Mobility data is incorporated to reflect the effects of human activities. As a result, the PDE model can be useful in making localized policies for reducing the spread of COVID-19.

We regroup 15 counties in Arizona into three regions: Northern Arizona, Central Arizona, and Southern Arizona. One may be able to use ordinary differential equations on multiple patches for the prediction of COVID-19 cases in Arizona. Here we choose partial differential equations instead of ordinary differential equations mainly because of the geographical and social characteristics in Arizona in terms of the spread of COVID-19. The three regions are geographically located in the north-south direction of Arizona and can be naturally projected onto the x-axis of the Cartesian coordinates. In addition, Central Arizona, home to the metro Phoenix area, has about 60\% of the population in Arizona.  As a result, the spread of COVID-19 between Central Arizona and other two regions are more likely predominant in terms of transboundary spread. These geographical characteristics, along with the disparities of social distancing and face mask requirements among the Arizona counties, fit squarely into the framework of reaction-diffusion equations for the spread of COVID-19. 

While capturing the trend of the spread of COVID-19, we were able to avoid large scale computation by regrouping the counties, which allowed us to use a computational-heavy PDE model.  Thus, the diffusion process of COVID-19 between county-regions as well as the increase of COVID-19 cases within a region are incorporated into the PDE model. The present paper aims to develop, for the first time in the literature, a specific PDE model between county-regions for forecasting daily COVID-19 cases and further quantifying the influences of the mobility of the community on the changing the number of COVID-19 cases in the state of Arizona. In addition,  In the present study, we study the effectiveness of human precautions such as face masks and social distancing on decreasing COVID-19 cases and hope to provide analytical results to help slow the spread of COVID-19 in Arizona.  

In this paper, we choose three regions of counties in Arizona for the spatial model because of the availability of both COVID-19 data and Google Community Mobility data. To gain more meaningful predictions with finer granularity, one may use zip codes or cities for the construction of regions or clusters. We could choose clustering algorithms to determine appropriate regions to ensure the connected zip codes stay as much as possible. However, the Google Community Mobility data is not available at the zip code level. Therefore, our model for prediction for regional COVID-19 is based on three regions of counties which is the most cost-effective method for prediction of COVID-19 in Arizona. 

Arizona has emerged as a COVID-19 hot spot since the stay-home orders were lifted in mid-May. It is important to study how to slow the virus’ spread. This study first presents a PDE model to describe  COVID-19  cases from community transmission, transboundary transmission between regions, and human activities. We apply this model for the prediction of COVID-19 cases to see the effect of social distancing on COVID-19 cases. The numerical results show that our new method of predicting COVID-19  cases is of acceptable high accuracy and can predict the effects of social distancing. To conclude, the prediction accuracy and the policy simulation suggest that a combining PDE model with real data can provide substantial information on COVID-19 spatio-temporal dynamics.

\section*{Acknowledgments}
The first author was supported by NSF grant \#1737861. 
\section*{Conflict of interest}
 The authors declare that they have no conflict of interest.



\end{document}